\documentclass[aps,prl,twocolumn,groupedaddress,maxnames=1]{revtex4-1}
\usepackage{graphicx}
\usepackage{natbib}
\usepackage{longtable}

\begin{document}


\title{Novel metastable metallic and semiconducting germaniums}



\author{Daniele Selli$^1$, Igor A. Baburin$^{1,2}$, Roman
  Marto\v{n}\'{a}k$^3$, Stefano Leoni$^{1,}$}
\email{stefano.leoni@chemie.tu-dresden.de} \affiliation{ 
$^1$Technische
  Universit\"at Dresden, Institut f\"ur Physikalische Chemie, 01062
  Dresden, Germany\\
  $^2$Max-Planck-Institut f\"ur Chemische Physik fester
  Stoffe, 01187 Dresden, Germany\\
  $^3$Department of Experimental Physics, Comenius University, Mlynsk\'{a}
  Dolina F2, 842 48 Bratislava, Slovakia}


\begin{abstract}
 By means of {\it{ab initio}} metadynamics runs we explored the lower-pressure region of the phase diagram of germanium. A monoclinic germanium phase with four-membered rings, less dense than diamond and compressible into $\beta$-tin phase (tI4) was found. A metallic bct-5 phase, mechanically stable down to room conditions appeared between diamond and tI4. mC16 is a narrow-gap semiconductor, while bct-5 is metallic and potentially still superconducting in the very low pressure range. This finding may help resolving outstanding experimental issues.
\end{abstract}


\maketitle




%

\section{Introduction}
The fundamental character and technological relevance of group-IVa elements (tetrels) have motivated repeated investigations and systematics on their polymorphism~\cite{Mujica:2003tb, Cui:2009kd, Chen:2011bx, Schwarz:2004fs, Katzke:2007cn}. Carbon polymorphs are promising as hard and transparent materials~\cite{Oganov:2006fu, Selli:2011kq}. Semiconducting silicon (Si) is versatile for micro- and nanoelectronic devices, while germanium (Ge) displays comparatively higher carrier mobility, a finer band gap tunability, and good compatibility with high-dielectric constant materials~\cite{Claeys:2010ij}. Metallization occurs in silicon and germanium upon compression~\cite{Mujica:2003tb}. In Ge lowering of phonon frequencies promotes electron-phonon coupling towards superconductivity~\cite{Chen:2011bx,Cui:2009kd}. The possibility of metallic germanium under room conditions is very intriguing and intensively debated~\cite{Cui:2009kd,Li:2010ge}, while superconductivity in elemental Ge appears under pressure~\cite{Chen:2011bx}. In the lower pressure range, improved optical properties via band-gap tuning can be achieved in a different polymorph.  

Engineering viable new compounds with superior properties entails a detailed understanding of structural changes~\cite{Oganov:2011vt}. Under pressure germanium bears similarities with silicon~\cite{Mujica:2003tb} by comparatively higher transition pressures with respect to Si, due to Ge core d-electrons~\cite{Lewis:1994tt}. Upon compression semiconducting Ge (cubic diamond) transforms into $\beta$-tin type (space group I4$_1$/amd) at about 10 GPa~\cite{Menoni:1986wv}, and then to Imma phase~\cite{Nelmes:1996ip}, simple hexagonal (P6/mmm)~\cite{Vohra:1986ge}, followed by orthorhombic Cmca phase~\cite{Takemura:2000tc} and finally upon further compression above 180 GPa, by the hexagonal close-packed arrangement (P6$_3$/mmc)~\cite{Takemura:2000tc}. 

The phase diagram of germanium is further complicated by a family of tetrahedral structures~\cite{Schwarz:2008wl,Guloy:2006gi}. Type-II clathrate Ge(cF136) exists at ambient conditions~\cite{Guloy:2006gi}. Other germanium modifications are reported, Ge(tP12), Ge(cI16) ($\gamma$-silicon type, BC8) and Ge(hR8)~\cite{Schwarz:2008wl}. BC8 Ge~\cite{Nelmes:1993va} is accessible through decompression from $\beta$-tin Ge. In a nutshell, during the last few years, new dense and open phases of germanium have been experimentally observed or synthesized. Nevertheless, a systematic approach to including known and finding novel germanium forms is still outstanding. 

\begin{figure}[t!]
\begin{center}
\includegraphics[width=0.50\textwidth,keepaspectratio]{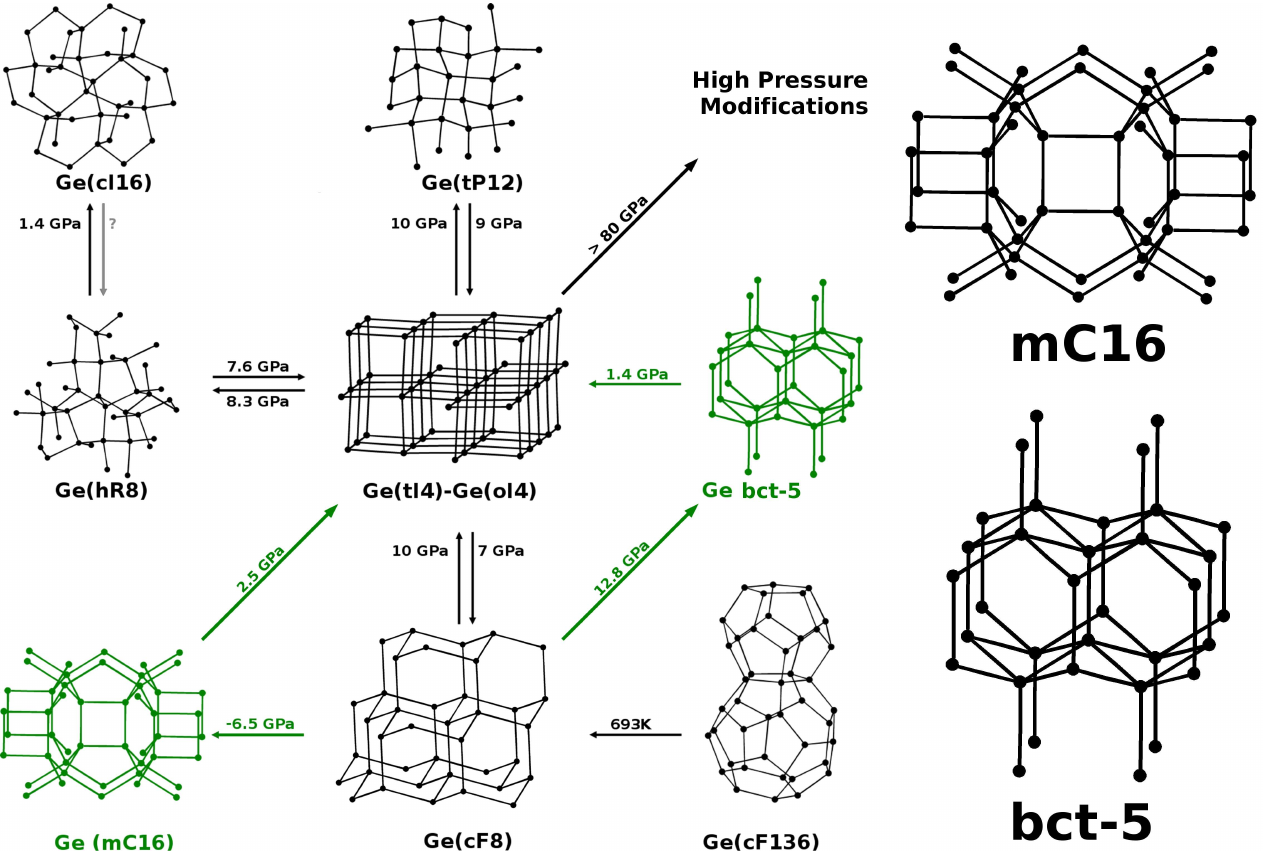}
\end{center}
\caption{Lower pressure region of the phase diagram of Ge, augmented by two novel phases mC16 and bct-5, found by {\it{ab initio}} metadynamics runs. bct-5 shows characterisitic square pyramidal 5-fold coordination of Ge atoms. In monoclinic mC16 four-rings are a characteristic feature. The arrows indicate the direction of metadynamics evolution. The pressures were evaluated based on the common tangent construction (see below, Fig.~\ref{Fig_2}).} 
\label{Fig_1}
\end{figure}

In this Letter we explore the energy landscape of germanium, both at ambient conditions and upon moderate compression. By means of {\it{ab initio}} metadynamics, we efficiently sampled structural transformations along appropriate collective reaction coordinates. Besides the already known dense phases of germanium, we found two novel allotropes (green in Fig.~\ref{Fig_1}). The first one is a monoclinic modification of germanium (mC16) slightly less dense than diamond, is an indirect band gap semiconductor and is unprecedented for tetrel elements. The second one is a five-coordinated (square pyramidal) metallic intermediate structure (tI4, sp. gr. I4/mmm), which incurs in the diamond (cF8) $\rightarrow$ $\beta$-tin phase (tI4) transition, and which has been postulated to exist in homologue silicon (bct-5)~\cite{Boyer:1991wb}. We report on transformation paths, energetic, mechanical and electronic properties. For the metallic bct-5 phase, we calculated superconducting temperatures down to ambient pressure compared to $\beta$-tin phase, based on the electron-phonon coupling mechanism~\cite{Baroni:2001tn}. 

\section{methods}
Different approaches are successful in the prediction of novel polymorphs
of the elements~\cite{Ma:2009et, Oganov:2009km}. Important discoveries have
been achieved by means of random techniques, genetic (evolutionary)
algorithms, or accelerated molecular dynamics~\cite{Oganov:2011vt}.
Metadynamics~\cite{epjb2011,p13,2006NatMa} allows for the exploration of
the energy surface along one or more collective reaction coordinates. The
method is independent of the level of theory used, it does not require
prior knowledge of the energy landscape and its sampling efficiency can be
enhanced by parallel runs started from different configurations. The
time-evolution of the system is biased by a history-dependent potential
constructed as a sum of Gaussians centered along the trajectory, in order
to discourage the system from visiting already harvested regions of the
potential~\cite{Laio:2008iw}.

All metadynamics runs were performed with at least eight atoms in the simulation box which served as a collective (6-dimensional) variable. This minimal box approach was successfully used in the prediction of novel carbon polymorphs, recently~\cite{Selli:2011kq}. The size of the minimal box ensured commensurability of all already known phases (except for the clathrate II phase, which requires a minimum of 34 atoms) either open or dense. Each metadynamics metastep consisted of molecular dynamics runs in the NVT ensemble for 0.5 ps (timestep 2 fs) at 300 K.

Metadynamics was performed with different molecular dynamics layers. A Density Functional Tight Binding (DFTB)~\cite{Frauenheim:2000tw} level of theory, as implemented in the $\Gamma$-point-only DFTB module of the CP2K code~\cite{p17,p18}, ensured rapid and accurate sampling in the low-pressure regime, characterized by four-connected Ge atoms.  For higher-pressures SIESTA~\cite{Soler:2002wq} was used as the DFT molecular dynamics layer, allowing for k-point runs. Electronic states were expanded by a single-$\zeta$ basis set constituted of numerical orbitals with a norm-conserving Troullier-Martins~\cite{Troullier:1991wi} pseudopotential description of the core levels. Single-$\zeta$ basis set dramatically reduces computational times providing nonetheless, the right topology and energy differences of all the Ge allotropes under study. The charge density was represented on a real-space grid with an energy cutoff ~\cite{Soler:2002wq} of 200 Ry. A Monkhorst-Pack k-point mesh of 2 $\times$ 2 $\times$ 2 ensured the convergence of the electronic part. High-pressure metadynamics was performed based on DFT. Lower pressure regions were initially explored by DFTB, followed by DFT metadynamics upon discovery of interesting novel polymorphs. In the lower pressure range the transferability between DFTB and DFT is unflawed.  


Electronic structure, phonon dispersion curves and superconducting properties were calculated with the Quantum Espresso (QE)~\cite{Giannozzi:2009hx, Baroni:2001tn} package. The superconducting critical temperature T$_c$ was evaluated based on the Allen and Dynes modification of the McMillan formula. This required calculating the electron-phonon coupling strength $\lambda$ via the Eliashberg function. The Coulomb potential value was $\mu$=0.1. A q-mesh of 8 $\times$ 8 $\times$ 8 was used for the evaluation of the dynamical matrix, while a Monkhorst-Pack k-point mesh of 12 $\times$ 12 $\times$ 12 ensured convergence of the electronic part.      

The structures visited during each run were characterized by their vertex
symbols, which contain the information on all the shortest rings meeting at
each atom, and coordination sequences, as implemented in the TOPOS
package~\cite{TOPOS}. In case of new structures ideal space group and
asymmetric units were identified with the Gavrog Systre
package~\cite{SYSTRE}. Subsequently a variable-cell geometry optimization
was performed (DFT-GGA, PBE functional~\cite{Perdew:1996ug}) in a plane-wave
pseudopotential framework~\cite{Giannozzi:2009hx} using Vanderbilt ultrasoft
pseudopotential as supplied by Perdew-Zunger with non-linear core
correction~\cite{Perdew:1981ug, Vanderbilt:1990ug}. A k-point mesh of 8 $\times$ 8 $\times$ 8
ensured convergence of the electronic part, while a plane-wave basis set with an
energy cut-off of 30 Ry was applied.


\section{results and discussion}
\begin{figure}[t!]
\begin{center}
\includegraphics[width=0.45\textwidth,keepaspectratio]{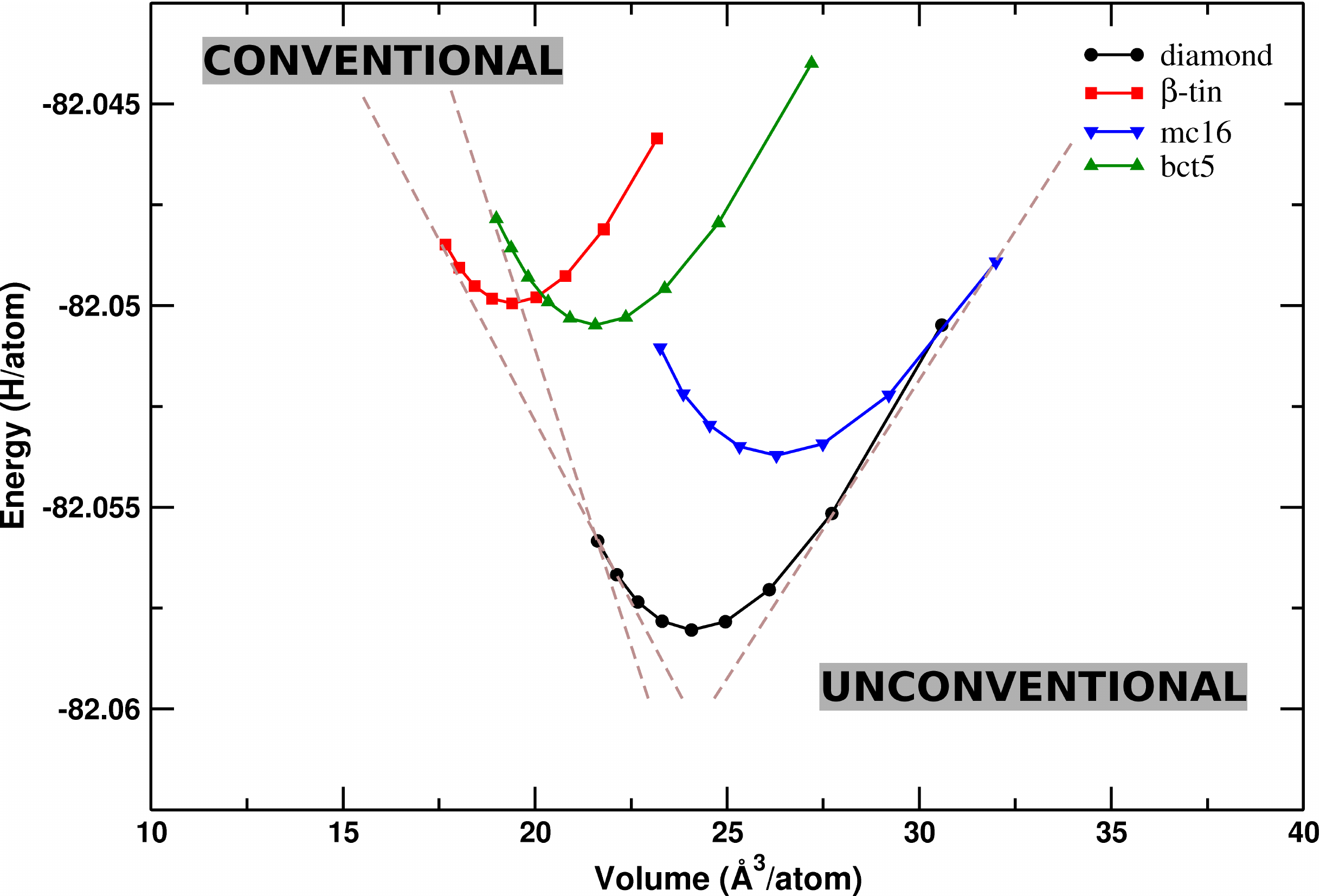}
\end{center}
\caption{Equation of states of Ge mC16 and bct-5, compared to cubic diamond and $\beta$-tin type. bct-5 features a reduced volume per atom compared to diamond type (cF8), while the total energy minimum lies lower than $\beta$-tin. mC16 on the contrary is less dense and energetically close to the diamond type. The tangents used for evaluating equilibrium pressures of Fig.~\ref{Fig_1} are highlighted.}
\label{Fig_2}
\end{figure}

The mC16 structure (Fig.~\ref{Fig_1}, C2/m, $a$=7.6094 \AA, $b$=7.9746 \AA,
$c$=6.5668 \AA, $\beta$=104.10$^{\circ}$) arised from a metadynamics run
started from diamond (8 atoms box, $p$=1 bar, $T$= 300 K). Ge atoms occupy
three Wyckoff positions: (4{\it{i}}) 0.70984 0.50000 0.67434, (4{\it{i}})
0.60981 0.50000 0.29080, (8{\it{j}}) 0.65012 0.76388 0.11596. Strikingly,
mC16 is less dense than diamond (see Fig.~\ref{Fig_2}), although
topologically as dense as diamond or lonsdaleite. Its bulk modulus amounts
to 51.2 GPa that is slightly lower than that of the diamond type-structure
(60.7 GPa), estimated from the fit to the third order Birch-Murnaghan
equation of state. Applying pressure to the mC16 allotrope in an additional
metadynamics run resulted into a direct transition to the $\beta$-tin
phase. Decompressing the latter is known to generate metastable phases
typically denser than diamond Ge. Therefore, a viable route to mC16, like
for other recent germanium allotropes, could rather be the oxidation of
suitable germanium Zintl salt precursors, i.e. via chemical synthesis.

Upon compression diamond transforms into $\beta$-tin and it subsequently
follows the same transition sequence of silicon phases. Along the diamond
$\rightarrow$ $\beta$-tin transition metadynamics (64 atoms box, $p$= 10
GPa, $T$= 300 K) visited an intermediate of bct-5 (sqp) topology (I4/mmm,
$a$=3.5491, $c$=6.4478, Ge(4{\it{e}}) 0.0 0.0 0.19273). The bct-5 bulk
modulus is 58.7 GPa, slightly lower than that of the $\beta$-tin phase
(68.2 GPa). This five-connected structure has been proposed for
silicon~\cite{Boyer:1991wb}, but has never been observed so far. The total
energy/volume curves of Fig.~\ref{Fig_2} suggest bct5 as a conventional
product of diamond compression. However, under hydrostatic conditions
$\beta$-tin is formed from diamond, while decompression leads to other
germaniums, although indications of minority phases exist.

\begin{figure}[b!]
\begin{center}
\includegraphics[width=0.45\textwidth,keepaspectratio]{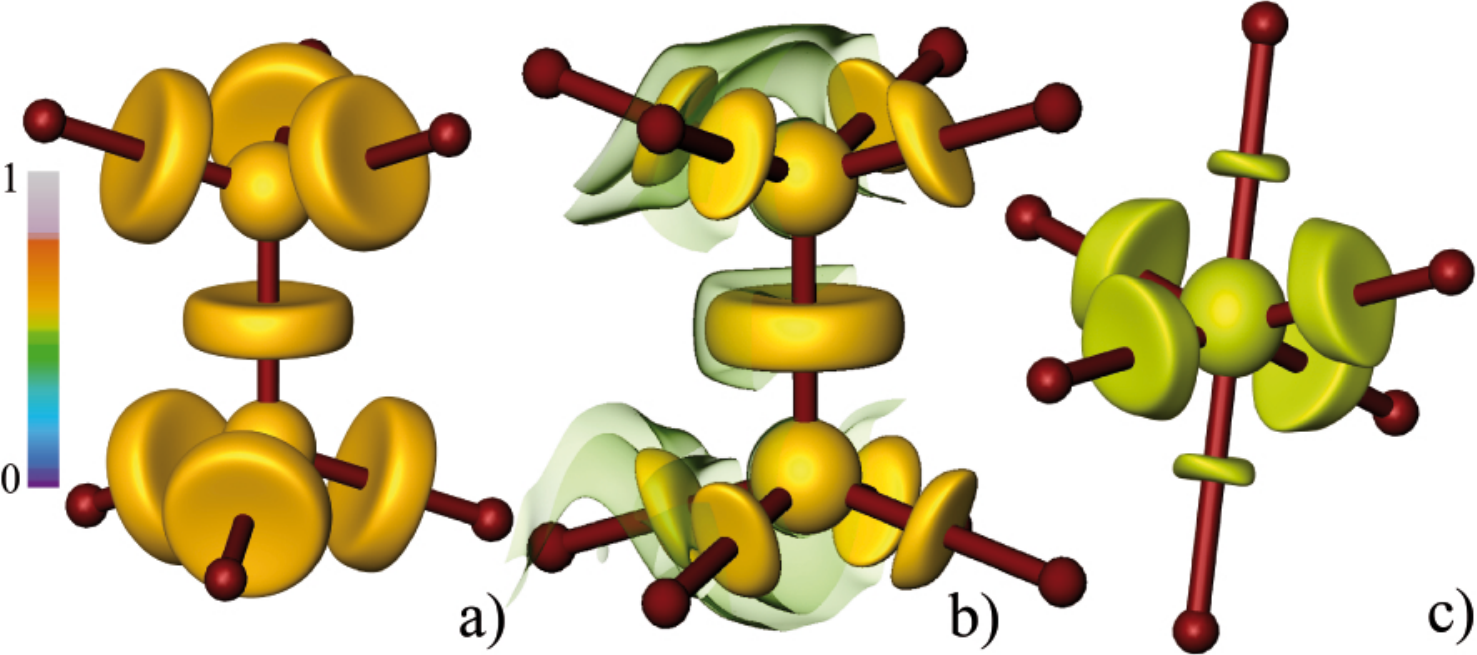}
\end{center}
\caption{Evolution of the bonding situation from diamond to $\beta$-tin type. The ELF map is showing four bond attractors for diamond Ge (a, $\eta=0.58$), one + four bond attractors for bct-5 (b, $\eta=0.53$, transparent green isosurface $\eta=0.48$), two + four bond attractors for $\beta$-tin (c, $\eta=0.51$).}
\label{Fig_3}
\end{figure}

\begin{figure}[t]
\includegraphics[width=0.48\textwidth,keepaspectratio]{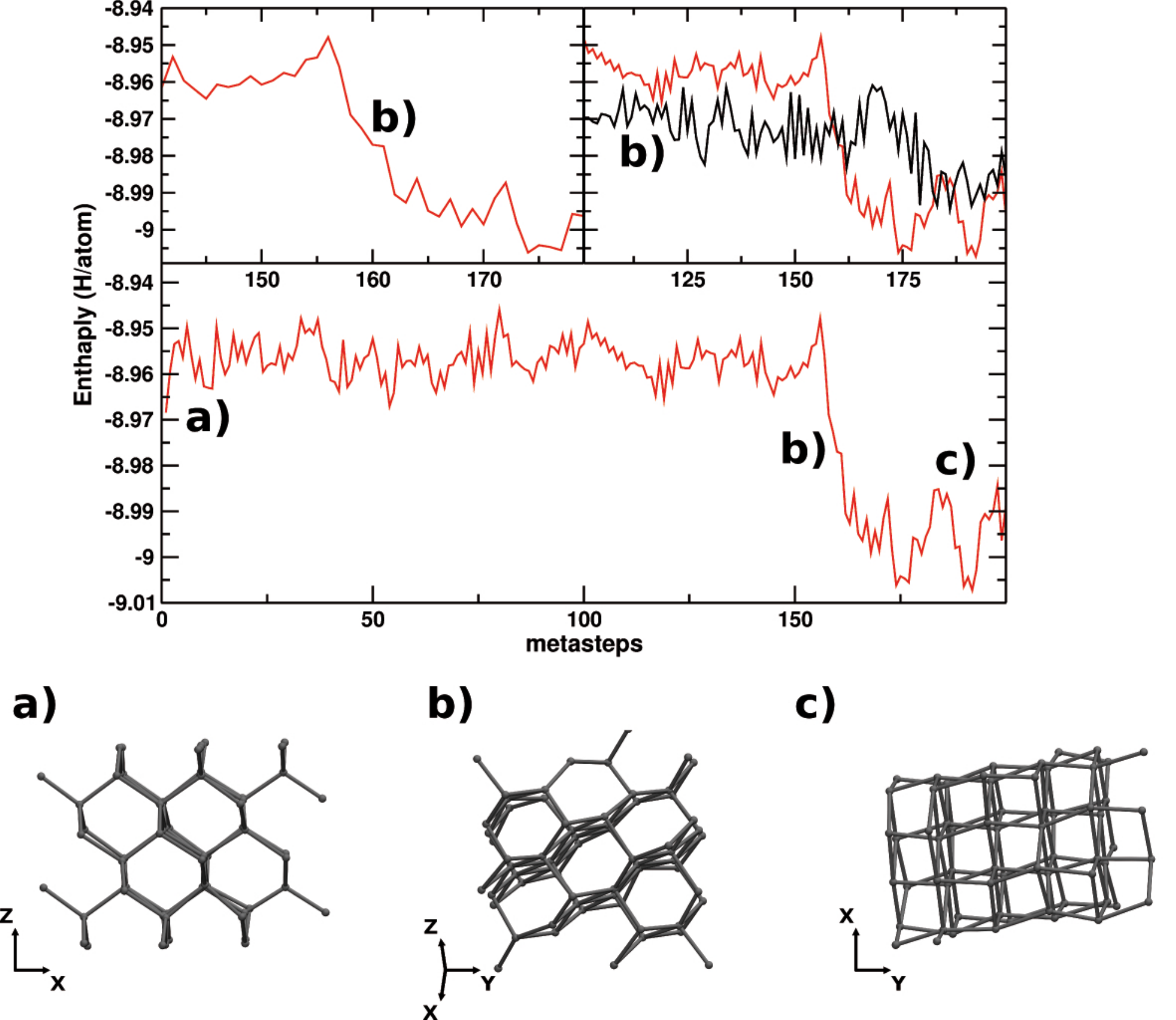}
\caption{Metadynamics (DFT) runs in a 64 Ge atoms box ($p_1$=10.0 GPa, T=300 K). bct-5 appears as a stepwise feature in the enthalpy profile of the run started from diamon (red line). Runs commenced from bct-5 (black line) evolve into $\beta$-tin. Configurations corresponding to distinct points along the runs are detailed below the graph.}
\label{Fig_4}
\end{figure}

The transformation affects only one box parameter, suggesting
nonhydrostatic shearing as the protocol of choice towards bct-5, also
supported by the magic-stress approach which led to bct-5 in silicon.
Alternatively, low-temperature compression may be considered. The evolution
of the enthalpy profile of metadynamics runs from diamond Ge
(Fig.~\ref{Fig_4}) shows bct-5 as a narrow plateau around metastep 160.
Metadynamics runs started from bct-5 (Fig.~\ref{Fig_4}, black curve)
confirms it as a proper intermediate along the transition, which can be
quenched down to room pressure, and which is mechanically stable (see
below). Mechanistically, the coordination number increases from 4 to 5 on
shortening one bond, followed by flattening of the pristine tetrahedron and
formation of the square pyramidal (sqp) geometry of bct-5,
Fig.~\ref{Fig_4}b. The four bonds in the pyramid basis are 2.62~\AA~long, the axial 2.48~\AA. 

The evolution of the bonding situation from Ge diamond to $\beta$-tin over bct-5 is shown in Fig.~\ref{Fig_3}. Calculation of the ELF~\cite{Becke:1990us} shows four, one + four and two + four bond attractors, respectively. The five ``bonds'' in this orbital-deficient, electron-deficient metallic bct-5 result from the $sp$ Ge valence shell. This bonding scenario is reminiscent of the recently discovered superconducting Zintl phase CaGe$_3$~\cite{Schnelle:2012kc}, isosymmetric with bct-5. 



\begin{figure}
\begin{center}
\includegraphics[width=0.40\textwidth,keepaspectratio]{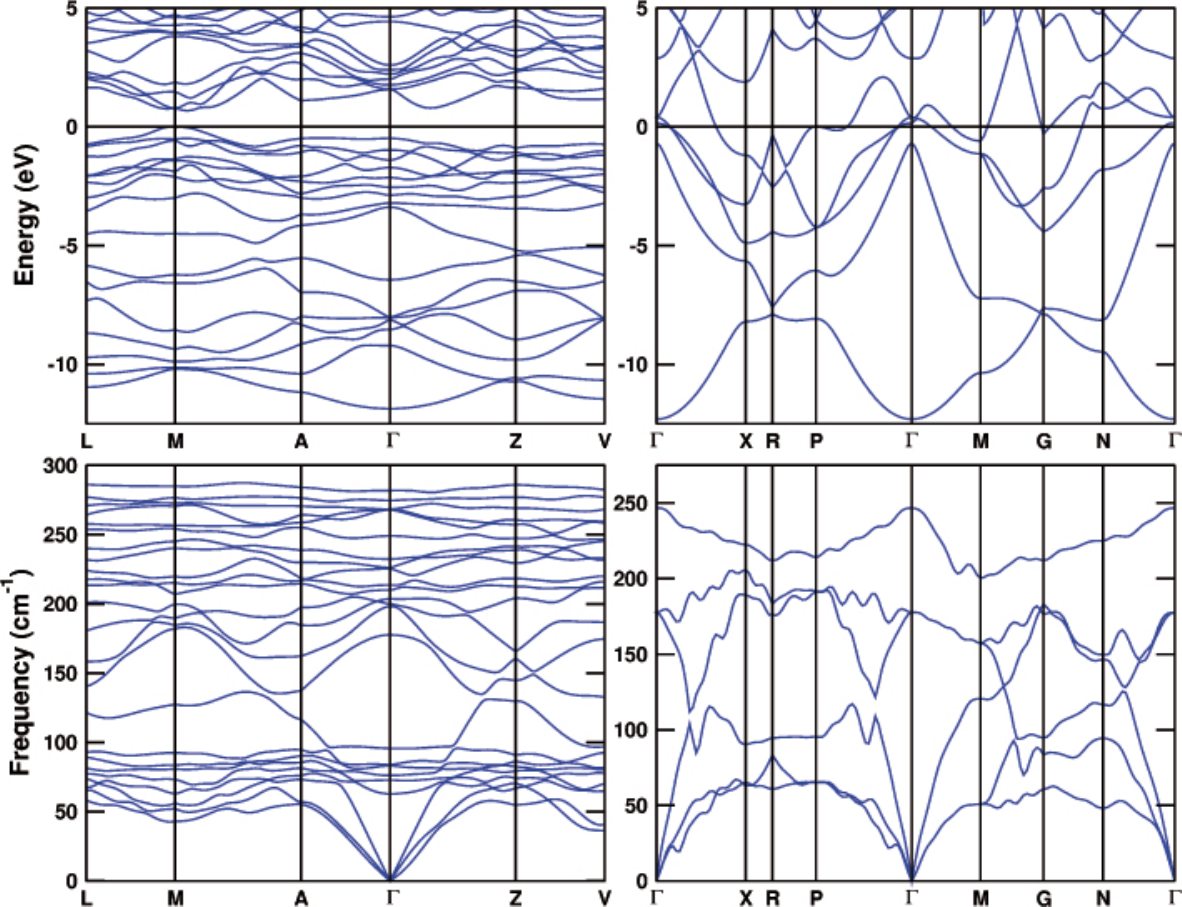}
\end{center}

\caption{Band structures and phonon spectra~\cite{Giannozzi:2009hx}(0.0 GPa) of mC16 (left) and bct-5. mC16 is a narrow-gap semiconductor (band gap = 1.43 eV), while bct-5 is metallic. Both are mechanically stable.}
\label{Fig_5}
\end{figure}

The electronic band structures and phonon dispersions of mC16 and bct-5 are shown in Fig.~\ref{Fig_5}. The tetrahedral phase is semiconducting with an indirect band gap of 1.43 eV (PBE-GGA~\cite{Perdew:1996ug}), while bct-5 is metallic and stable down to 0 GPa. Isothermic-Isobaric molecular dynamics (1 bar, 300 K, 2.5 ps) confirmed the stability of bct-5. mC16 is characteristic due to the presence of four-rings. However, this does not imply overall structure destabilization~\cite{Karttunen:2011en}. The expectation of a strained geometry is in fact not supported by total energy calculations, which place mC16 among the energetically lowest Ge allotropes. The indirect band gap and the low density (compared to the diamond type) makes this germanium an attractive material. The need for a ``negative'' pressure makes a chemical path plausible.

\begin{figure}[t!]
\begin{center}
\includegraphics[width=0.40\textwidth,keepaspectratio]{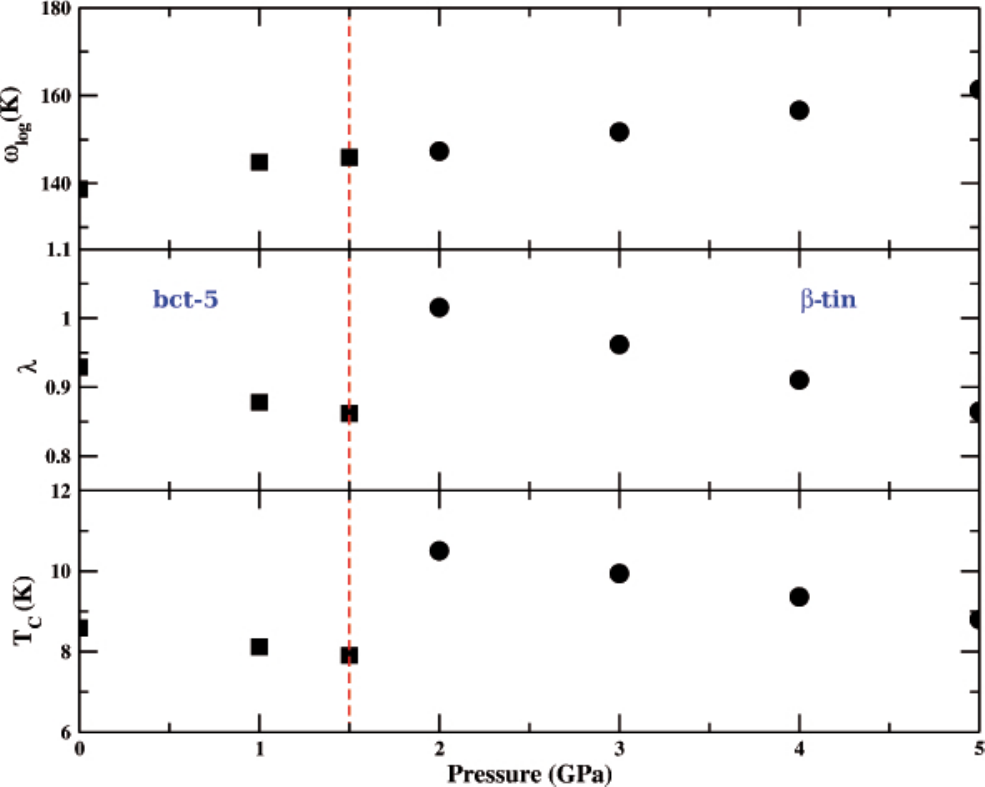}
\end{center}
\caption{Evolution of $\omega$, $\lambda$ and T$_c$ as a function of pressure for bct-5 and $\beta$-tin, calculated based on the electron-phonon coupling model. The calculated equilibrium pressure between bct-5 and $\beta$-tin marks the boundary between the phases. The model predicts an increase of T$_c$ in the lower pressure region.}
\label{Fig_6}
\end{figure}

The prominent property of bct-5 is the conservation of metallic character down to ambient conditions. Calculations and experiments have shown an increase of the superconducting temperature on lowering pressure, with superconductivity still present around 8,7 Gpa (T$_c$= 5 K)~\cite{Cui:2009kd}. The evolution of T$_{c}$ as a function of pressure for bct-5 is shown in Fig.~\ref{Fig_6}. The ongoing debate on the possibility of ``strange metallic'' states other than $\beta$-tin~\cite{Cui:2009kd,Li:2010ge}, and the need for further explanations of the survival upon decompression of certain ``metallic'' metastable states that could not be reliably assigned to any known Ge phase, open the door for a serious consideration of the role of the bct-5 in the lower pressure range, as a metallic state that qualifies for higher T$_c$ values (via the McMillan relation).

In conclusion, with metadynamics runs we have found and characterized two
novel Ge polymorphs. The first one, mC16, is an indirect gap semiconductor, and
its structure is unprecedented for the tetrel family. The second bct-5
polymorph was suggested for Si. Our simulations lean strong relevance to
bct-5 in the lower pressure range, as a further metallic superconducting phase capable of
stability at room conditions. We expect our predictions to stimulate
further experimental work.

\begin{acknowledgments}
  R.M. was supported by the Slovak Research and Development Agency
  under Contract No. APVV-0558-10 and by the project implementation
  26220220004 within the Research \& Development Operational Programme
  funded by the ERDF. S.L. thanks the DFG for support under the priority
  project SPP 1415, as well as ZIH Dresden for the generous allocation of
  computational resources. We acknowledge discussions with K. Syassen and
  E. Tosatti.
\end{acknowledgments}

\bibliographystyle{apsrev4-1}
\bibliography{ge.bib}

\end{document}